\documentclass[submission,copyright,creativecommons,noderivs,colorlinks]{eptcs}
\providecommand{\event}{SYNT 2014} 
\usepackage{breakurl}             
\usepackage{amssymb}
\usepackage{amsmath}
\usepackage{tikz}
\usetikzlibrary{patterns}
\newtheorem{example}{Example}
 
\title{Low-Effort Specification Debugging and Analysis}
\author{R\"udiger Ehlers \institute{University of Bremen \& DFKI GmbH \\ Bremen, Germany} \and Vasumathi Raman
\institute{California Institute of Technology\\Pasadena, CA, United States}
}
\def\titlerunning{Report-Based Specification Analysis and Debugging}
\def\authorrunning{R\"udiger Ehlers \& Vasumathi Raman}

\newcommand{\newterm}{\emph}
\newcommand{\markterm}{\relax}
\DeclareMathOperator{\LTLG}{\mathsf{G}}
\DeclareMathOperator{\LTLX}{\mathsf{X}}
\DeclareMathOperator{\LTLF}{\mathsf{F}}
\DeclareMathOperator{\TRUE}{\mathbf{true}}
\DeclareMathOperator{\FALSE}{\mathbf{false}}
\newcommand{\NN}{\mathbb{N}}
\newcommand{\AP}{\mathsf{AP}}

\begin{document}
\maketitle

\begin{abstract}
Reactive synthesis deals with the automated construction of implementations of reactive systems from their specifications. To make the approach feasible in practice, systems engineers need effective and efficient means of debugging these specifications. 
In this paper, we provide techniques for \emph{report-based} specification debugging, wherein salient properties of a specification are analyzed, and the result presented to the user in the form of a report. This provides a low-effort way to debug specifications, complementing high-effort techniques including the simulation of synthesized implementations.
We demonstrate the usefulness of our report-based specification debugging toolkit by providing examples in the context of generalized reactivity(1) synthesis.
\end{abstract}

\section{Introduction}
\label{sec:intro}
A modern approach to the engineering of correct-by-construction reactive systems is \newterm{synthesis}, wherein a reactive controller is constructed automatically from its specification, which it is guaranteed to satisfy.
Recent work in robotics and control has shown the applicability of the approach to the construction of a wide variety of systems \cite{KGF,Nok10,Nuzzo13}. 
The most commonly used synthesis workflow in this context has been \newterm{generalized reactivity(1) synthesis} \cite{DBLP:journals/jcss/BloemJPPS12}, in which the full expressivity of a temporal logic such as linear-time temporal logic (LTL) \cite{DBLP:conf/focs/Pnueli77} is traded against the existence of a more efficient symbolic synthesis algorithm (in this case for a subset of LTL). 

As appealing as the concept of reactive system synthesis may be, it merely reformulates the challenge of constructing a reactive system from \emph{designing the right system} to \emph{designing the right specification}. 
While a specification must arguably always be written in order to verify the correctness of a manually engineered system, designing such a specification is usually harder in the context of synthesis. This is because, when  verifying a system that has already been constructed, the system and its specification can be cross-checked: whenever an example trace of the system is found that does not satisfy the specification, the trace can be manually inspected for whether it constitutes an error in the system or in the specification.
Thus, both the specification and the system can be iteratively refined during the verification process.
This is not an available option for synthesis, as there is no system to begin with, and so other means for ensuring correctness of the specification are needed.

This motivates a systematic \newterm{specification debugging} process. Starting from a \newterm{candidate specification}, the objective is to analyze the specification for missing or incorrect parts, and to correct it as needed. Corrections should usually only be done manually, as they need to capture the designer's intent, but algorithmic support is often valuable for the analysis.
Debugging techniques can be divided into two categories: 
\begin{enumerate}
\item those that find reasons for the \newterm{unrealizability} of a candidate specification when the specification is over-constrained and no implementation exists, and
\item those that analyze \newterm{realizable specifications}, and try to find missing \newterm{guarantees} that the synthesized system needs to fulfill, or overly restrictive \newterm{assumptions} about the operating environment for the synthesized system.
\end{enumerate}
The classical approach to debugging a specification is by \emph{game-based simulation} \cite{DBLP:journals/sttt/KonighoferHB13,DBLP:journals/taosd/MaozS13}. For a realizable specification, a simulation of the synthesized system is offered, in which the user can provide inputs to the system and observe its reactive behavior. For an unrealizable specification, the roles of the user and the simulation are exchanged: the user is asked to simulate the system, and the simulated environment provides inputs to the system that force it to violate its specification.

Game-based simulation can be considered a ``high-effort'' approach to specification debugging: it requires the user to make choices during the simulation that identify missing environment assumptions or system guarantees, and to keep track of system and environment obligations. 
Particularly for systems with many input and output variables, choosing next suitable actions is difficult and error-prone. 
Interpreting the simulated moves in the game can be equally difficult, requiring domain-specific interpretation and visualization support, which imposes a substantial implementation workload.
Even custom application support does not guarantee that the opponent's behavior in the simulation game is easy to understand, especially when debugging unrealizable infinite-horizon goal or liveness specifications.

These observations motivate augmenting the specification designer's toolbox with low-effort techniques that reduce the number of scenarios in which a game-based simulation is required. 
Some such techniques are practically folklore: e.g., it is easy to check whether an environment assumption is needed, by removing it and checking that the specification remains realizable.
Superfluous environment assumptions indicate potential problems with the specification, such as a poorly formulated system guarantee whose fulfillment the assumption was originally introduced to ensure.

Despite the existence of these folklore approaches, a systematic and more fine-grained account of low-effort specification debugging techniques is, to the best of our knowledge, still missing in the synthesis literature. Our main contribution in this paper is a comprehensive set of tools for low-effort specification debugging. All of our techniques are \emph{report-based}, i.e., they analyze a specification and produce a \emph{report} identifying potential problems. All components of our tool set are push-button techniques, allowing a one-step analysis of a specification. We discuss several folklore techniques as well as new approaches, and extend the former to more fine-grained versions where applicable. For example, the check for superfluity of an assumption can be refined by asking whether the assumption can help the controller achieve its goals sooner.

Our techniques bolster the main premise of reactive synthesis: they enable quick and efficient construction of reactive systems from their specifications, without needing to construct a system from scratch when the specification changes, and with minimum effort for the system engineer. 
In being report-based, our techniques allow the easy detection of specification errors by the user, since the generated report can simply be checked for unexpected results. 
While our techniques certainly do not identify \emph{all} problems with a specification (e.g. we cannot detect the incorrect interpretation of a system's objective by an engineer), they substantially contribute to the appeal of reactive synthesis as an integral part of a system's engineer's toolkit.

All of the techniques presented in this paper have been implemented for generalized reactivity(1) specifications using the generalized reactivity(1) synthesis tool \textsc{slugs}. 
Some of the techniques are implemented as scripts that simply call \textsc{slugs}, while others are implemented as plugins for \textsc{slugs}.
We also provide examples of the errors that can be discovered by applying the proposed techniques.

\subsection{Related Work}

There is considerable prior work on debugging LTL formulas in the formal methods literature, and many of these approaches take a report-based perspective. For example, the problem of identifying small causes of failure has been studied from multiple perspectives. For unsatisfiable LTL formulas, the authors of \cite{Schuppan09} suggest several notions of unsatisfiable cores, with corresponding methods of extraction. Of these, the technique of extracting an unsatisfiable core from a Bounded Model Checking (BMC) resolution proof was used in \cite{Shlyakhter03} for debugging declarative specifications. There, the abstract syntax tree (AST) of an inconsistent specification was translated to CNF, an unsatisfiable core extracted from the CNF, and the result mapped back to the relevant parts of the AST. The authors of \cite{Cimatti07,Raman_IROS13} also attempted to generalize the idea of unsatisfiable and unrealizable cores to the case of temporal logic using SAT-based bounded model checkers.
Complementary to these techniques is the detection of \newterm{inherent vacuity} \cite{DBLP:conf/hvc/FismanKSV08} in a specification, which intuitively means that a part of the specification does not have any effect on its overall satisfaction by any system. Such cases are typically undesirable, and a common consequence of errors in the specification.  

In the context of unrealizability, the authors of \cite{Cimatti08} define a notion of ``helpful'' assumptions and guarantees, and compute minimal explanations of unrealizability by iteratively expelling unhelpful constraints. They assume an external black-box realizability checker for iterated realizability tests. The authors in \cite{DBLP:conf/hvc/KonighoferHB10} use model-based diagnosis to remove guarantees as well as irrelevant output signals from the specification; these are output signals that can be set arbitrarily without affecting the unrealizability of the specification. To identify and eliminate the source of unrealizability, there are also efforts such as those in \cite{Li11,Chatterjee08} that provide additional environment assumptions to make the specification realizable. This is accomplished in \cite{Chatterjee08} using efficient analysis of turn-based probabilistic games, and in \cite{Li11} by template-based mining of the environment counterstrategy.  The techniques we present in this paper have a spirit similar to the work in \cite{Cimatti08}, but we consider a larger variety of interesting properties of the specification to include in our reports.
 
In the context of game-based debugging, \cite{DBLP:journals/sttt/KonighoferHB13,DBLP:conf/hvc/KonighoferHB10} present a game for analyzing unrealizable specifications, and a tool implementation. 
The authors of \cite{DBLP:journals/taosd/MaozS13} also present a game-based approach. As elaborated in Section \ref{sec:intro}, we instead focus on a \emph{report-based} methodology.

The utility of both report-based analysis and game-based debugging was explored in the context of robotics by \cite{Raman_TRO13,Raman_CAV11}. These works provide domain-specific information about unrealizable specifications for high-level robot control. The ideas presented in this paper on the other hand are quite general, and not tailored to any particular application domain.

\subsection{Contributions}
The report-based specification debugging techniques presented in this paper are divided into four categories:
\begin{enumerate}
\item Preventing the accidental writing of unintended specifications
\item Analyzing the interaction of the \newterm{assumptions} and \newterm{guarantees} in the specification
\item Analyzing the influence of the individual input and output signals on the realizability of a specification
\item Generating concise ``informative'' strategies or traces of the synthesized system
\end{enumerate}
After reviewing some preliminaries on generalized reactivity(1) synthesis in Section \ref{prelims}, the contributions in each of the above categories are presented in Sections \ref{sec:techSpecWriting}, \ref{sec:assumptionsAndGuarantees}, \ref{sec:weakeningAndStrengthening}, and \ref{sec:strategiesAndTraces}, respectively. For most techniques, we present examples to provide the reader with an intuition for how they help with specification debugging. 
In order to be as illustrative as possible, all examples have intentionally been kept as simple as possible. Thus, readers who are experienced specification designers will likely identify the problems with the example specifications without the debugging techniques presented in the paper. Nevertheless, such problems are much more difficult to observe in larger specifications (which the presented examples could be contained within). Therefore, the techniques presented remain useful even for engineers with a lot of specification engineering experience.

We have implemented the proposed techniques in the generalized reactivity(1) synthesis tool \textsc{slugs}. This tool is open source and freely available for download at \url{http://github.com/ltlmop/slugs}. Versions later than \textit{May 3, 2014} feature 
the script \texttt{tools/createSpecificationReport.py} 
that runs all the proposed analyses in one step and outputs an HTML report of the results.

\section{Preliminaries} \label{prelims}

\paragraph{Words and Linear Temporal Logic:} Given an \newterm{alphabet} $\Sigma$, a \newterm{word} $w = w_0 w_1 \ldots$ is defined as a finite or infinite sequence of \newterm{letters} $w_i \in \Sigma$. For the rest of this paper, we will typically use $\Sigma = 2^{\AP}$ for some set of \newterm{atomic propositions} $\AP$. We can characterize subsets of such words by a formula in \newterm{linear temporal logic} (LTL). We only consider a sub-fragment of LTL here, where the formulas are built using the following grammar:
\begin{equation*}
\varphi = p \in \AP\, ||\, \neg \varphi\, || \, \varphi' \vee \varphi''\, || \, \varphi' \wedge \varphi''\, || \, \varphi' \rightarrow \varphi''\, ||\, \LTLX \varphi'\,||\,\LTLG \varphi'\,||\,\LTLF \varphi'
\end{equation*}
For semantics and a more complete description, we refer the reader to \cite{DBLP:conf/focs/Pnueli77}.

\paragraph{Finite-State Machines and Synthesis:}
The importance of LTL in the context of this paper is that it can describe the desired behavior of the system to be synthesized. Formally, we represent such systems as \newterm{finite-state (Mealy) machines} $\mathcal{M} = (S,\Sigma_I,\Sigma_O,s_0,\delta)$ with the set of states $S$, the input alphabet $\Sigma_I$, the output alphabet $\Sigma_O$, the initial state $s_0$, and the transition function $\delta : S \times \Sigma_I \rightarrow S \times \Sigma_O$. The finite-state machine (FSM) produces infinite traces $w = w_1 w_2 \ldots$, such that $w_i \in \Sigma_I \times \Sigma_O$ for every $i \in \mathbb{N}$. The FSM has a unique trace for every input sequence $w_I \in (\Sigma_I)^\omega$.

In the context of synthesis, we typically have $\Sigma_I = 2^{\AP_I}$ for a set of atomic input propositions $\AP_I$ and $\Sigma_O = 2^{\AP_O}$ for a set of atomic output propositions $\AP_O$. An LTL formula over $\AP_I \uplus \AP_O$ can describe properties to be satisfied by all executions (or traces) of a finite-state machine with \newterm{signature} $(\Sigma_I,\Sigma_O)$.

The \newterm{realizability} problem is to determine, given $\AP_I$, $\AP_O$ and a temporal logic specification $\psi$ over $\AP_I \uplus \AP_O$, whether there exists a finite-state machine $\mathcal{M} = (S,\Sigma_I,\Sigma_O,s_0,\delta)$, all of whose traces satisfy $\psi$. The \newterm{synthesis problem} additionally asks for the construction of such a system if one exists.

\paragraph{Generalized Reactivity(1) Specifications and Synthesis:} The synthesis problem for full LTL has been shown to have a relatively high complexity (2EXPTIME in the size of the formula \cite{DBLP:conf/icalp/PnueliR89}). To side-step this prohibitive complexity, generalized reactivity(1) \cite{DBLP:journals/jcss/BloemJPPS12} has emerged as a fragment of LTL with only singly-exponential time complexity of synthesis (see, e.g., \cite{Ehlers2013}). 
In this approach, which is commonly abbreviated as \newterm{GR(1) synthesis}, the allowed system specifications are restricted to the following form:
\begin{equation}
\label{eqn:GR1Equation}
( \varphi^a_i \wedge \varphi^a_s \wedge \varphi^a_l ) \rightarrow_b 
( \varphi^g_i \wedge \varphi^g_s \wedge \varphi^g_l )
\end{equation}
The specification in Equation~\ref{eqn:GR1Equation} has two \newterm{components} separated by an implication operator. The two components are in turn conjunctions of sub-formulas in LTL, which we will call \newterm{specification parts} for the scope of this paper. The parts left of the implication operator are called the \newterm{assumptions}, whereas the parts right of the implication operator are the \newterm{guarantees}. The implication operator itself represents a \newterm{strict implication}. As defined by Bloem et al.~\cite{DBLP:journals/jcss/BloemJPPS12}, this intuitively means that a system that satisfies the specification must ensure that the right-hand side of the formula is not violated before the left-hand side is violated.
The specification parts $\varphi^a_i$, $\varphi^a_s$, $\varphi^a_l$, $\varphi^g_i$, $\varphi^g_s$, and $\varphi^g_l$ also have a specific form:

\begin{itemize}
\item $\varphi^a_i$ and $\varphi^g_i$ are conjunctions of initialization assumptions and guarantees, each of which are free from temporal operators. In addition, no proposition from $\AP_O$ can be used in $\varphi^a_i$.
\item $\varphi^a_s$ and $\varphi^g_s$ are conjunctions of safety assumptions and guarantees. Each such conjunct is a sub-property represented by a temporal logic formula $\LTLG \psi$, where the only temporal operator occurring in $\psi$ can be $\LTLX$, and no nesting of temporal operators is allowed in $\psi$. Moreover, no proposition from $\AP_O$ may be used in the scope of an $\LTLX$ operator in $\varphi^a_s$.
\item $\varphi^a_l$ and $\varphi^g_l$ are conjunctions of liveness properties of the form $\LTLG \LTLF \psi$, in which the only temporal operators occurring in $\psi$ can be unnested occurrences of $\LTLX$.\end{itemize}

In contrast to some previous works that apply GR(1) synthesis, we allow the use of the next-time operator in liveness properties. It has been shown that this extension can be handled without changing the synthesis algorithm significantly \cite{DBLP:conf/icra/RamanPK13}.

As a further extension, we will sometimes refer to the so-called \newterm{robotics semantics} of GR(1) synthesis. This means that the system must be able to start with every possible output satisfying $\varphi^g_i$. In the robotics context, where the robotic system can control its position in the workspace, this usually corresponds to the requirement that the robot is able to start from any admissible workspace position. 

While the set of types of properties that are directly supported by GR(1) synthesis is relatively limited, some more complex properties can be encoded by performing \emph{pre-synthesis} \cite{DBLP:conf/fmcad/SohailS09,Ehlers2013}, where essentially the system is asked to output a certificate proving that it satisfies these more complex properties. Constraints that enforce that the synthesized system outputs a correct certificate can then be encoded in the form of GR(1) guarantees (see, e.g., \cite{Ehlers2013}).

If a given specification is \newterm{realizable}, i.e., there exists an implementation, the GR(1) synthesis algorithm produces an implementation of a specific form, where all states are labeled by the last input character seen $x \in \Sigma_I$, the last output character $y \in \Sigma_O$, and the current \newterm{goal} $\psi$ of the system. We say that an expression $\psi$ is a goal of the system if $\LTLG \LTLF \psi$ is a liveness guarantee in the specification. There is at most one state for each such labeling, imposing an upper bound on the number of states in the implementation.
 
In GR(1) specifications, liveness assumptions are introduced to allow the synthesized system to wait for certain input events that are guaranteed to be eventually triggered. Given an $(x,y,\psi)$-labeled state in a finite-state machine, the \newterm{reactive distance} of the state is the maximum number of waiting phases and transitions between waiting phases necessary for \emph{any} finite-state machine to enforce that some transition satisfying $\psi$ is eventually taken when starting from that state\label{secWithDef:reactiveDistance}. In GR(1) synthesis, the reactive distance of a state is computed before an implementation is built. In particular, a \newterm{synthesis game} is built from the specification, in which a winning strategy for one of the players in the game, the \newterm{system player}, represents an implementation for the system. The reactive distance of a state is then obtained as a by-product when \newterm{solving the game} to find this winning strategy.

Generalized reactivity(1) synthesis tools are typically implemented using \newterm{binary decision diagrams} (\newterm{BDD}s) as the symbolic data structure for manipulating position sets in the synthesis game. A BDD is a directed acyclic graph with nodes representing boolean variables. Every node has a $\FALSE$-successor and a $\TRUE$-successor, and there is additionally a $\FALSE$ sink and a $\TRUE$ sink in the graph. A BDD represents a boolean function, mapping variable valuations that induce paths leading to the $\TRUE$ sink to $\TRUE$. We call a partial assignment to some set of variables a \newterm{cube} for some boolean function over the same set of variables if the function maps all assignments that are concretizations of the partial assignment to $\TRUE$.

\section{Facilitating Specification Writing}
\label{sec:techSpecWriting}

The simplest errors in specifications are those that stem from a mismatch of the designer's intent with the specification that they actually write down. Such a mismatch can have a variety of causes, starting from simple typos, to an erroneous manual encoding of complex value domains into boolean signal values, to a misunderstanding of the semantics of the specification formalism.
To counter such specification writing errors, we propose two techniques:
\begin{itemize}
\item User support for more complex value domains in the specification language
\item Analyzing whether the GR(1) strict implication semantics affects the specification at hand.
\end{itemize}
The first of these techniques is not strictly debugging-specific, as its purpose is to allow representing the specification in a more concise and intuitive way. Thus, it \emph{reduces} the risk of introducing errors rather than \emph{detecting} them; we include it in this paper because it shares the goal of eliminating errors in specifications. 
Moreover, it complements the other specification debugging approaches presented, by allowing information about states in the synthesis game or traces of the system to be represented in a more human-readable form.

\subsection{Supporting a Richer Variable Value Domain}
Classically, GR(1) specifications govern purely boolean predicates, and there is no support for richer variable domains. There are many applications in which \markterm{variables} from a richer domain naturally occur; it is thus necessary to encode their values into boolean signal \markterm{valuations}. 
For example, in \cite{DBLP:conf/icra/Kress-GazitFP07,DBLP:conf/aaai/FinucaneJK11}, robot location regions are binary-encoded in order to reason about the location of the robot. Integer numbers are also a common occurrence in many specifications (see, e.g., \cite{EhlersTopcuHSCC2014,Wongpiromsarn2011-infotech,DBLP:conf/iccps/OzayTMW11}). This calls for user support in order to specify such scenarios concisely, without needing to perform a manual boolean encoding. Even if the encoding is already automated at the application level (as in \cite{DBLP:conf/icra/Kress-GazitFP07,DBLP:conf/aaai/FinucaneJK11}), leaving the encoding to the synthesis tool-chain enables debugging the specifications at a higher level.
Many tools for verifying systems, such as \textsc{NuSMV} \cite{DBLP:conf/cav/CimattiCGGPRST02} and \textsc{UPPAAL} \cite{DBLP:conf/movep/AmnellBBDDFHJLMPWY00} offer direct support for richer variable domains. For synthesizing systems, fewer software packages (e.g. \textsc{TuLiP} \cite{DBLP:conf/hybrid/WongpiromsarnTOXM11}) support integers as data types.

Our tool \textsc{slugs} supports the direct use of integer variables in specifications. All such variables have a lower bound $b_l$ and an upper bound $b_u$ (such that $b_u \geq b_l$). We use a preprocessing script for translating specifications with integer variables to boolean ones. This allows to treat the specification as purely boolean for most specification analysis steps. The preprocessor allocates precisely $\lceil \log_2( b_u-b_l+1 ) \rceil$ bits for an integer variable with bounds $[b_l,b_u]$. For integer variables that are an input to the system, initialization and safety assumptions are added that prevent the environment from picking a number that is not in the range $[b_l,b_u]$, and corresponding guarantees are added for output variables. 

Merely adding such support is however not quite enough to facilitate the writing of correct specifications. In order to allow operations such as integer addition, and to encode the synthesis game using BDDs as contemporary synthesis tools do, we have to define how to handle under- and over-flows of these integer ranges.
Consider as an example the property $\phi = \LTLG(a + b < 7)$, where $a \in \{0, \ldots, 7\}$ is an input to the system and $b \in \{0, \ldots, 7\}$ is an output of the system. If $\phi$ is a system guarantee, then it is easy to fulfill if the value of $a+b$ wraps above $7$, as the system can simply choose any $b \in \{1, \ldots, 7\}$ when $a = 7$, and $b=0$ otherwise. If $a$ and $b$ have different variable domains, things become even more complicated. Alternatively, we could use saturation semantics for $a+b$, or just declare that an equation resolves to $\mathbf{false}$ if either side exceeds the variable bounds. This would unfortunately lead to the inconvenient fact that $a + b < 7$ is no longer equivalent to $(a + b +i) < (7 + i)$ for every $i \in \NN$. The system engineer would need to consider the value domains of all variables at every step of writing the specification in order to avoid such unintended effects.

As a far simpler alternative, we propose to apply the semantics introduced in \cite{DBLP:conf/cav/PeterEM11}, where \textbf{expressions never overflow}, because all sub-expressions have $\NN$ as the range of variables.
Using these semantics, the expression $a + b +i < 7 + i$ is equivalent for all values of $i \in \NN$, and having $\phi = \LTLG(a + b < 7)$ as a guarantee in a specification leads to its unrealizability if $a$ is an input signal.
Accommodating this semantics in a BDD-based synthesis workflow can be done by computing an upper bound on the number of bits needed for storing the result of each sub-expression. As this number is always finite, a finite number of bits suffices. This information is then used to bound the number of BDDs to be computed when building the transition relation of the synthesis game.

\subsection{Strict Implication Semantics Analysis}

In GR(1) synthesis, the system to be synthesized may only violate some safety guarantee \emph{after} an assumption has been violated. Thus, the implication operator that separates assumptions and guarantees in GR(1) synthesis has a non-standard semantics. This difference is of importance when encoding more complex properties into GR(1) form, as it prevents the system from actively violating the guarantees in order to enforce the environment assumptions to be violated later. 
As an example, consider a GR(1) specification of the following form over $\AP_I = \{ p, q \}$ and $r \in \AP_O$:
\begin{equation*}
\psi = \left(
\LTLG( q \vee \LTLX q)
\,\wedge\,
\LTLG( \neg q \vee r)
\right) 
\,\rightarrow_S\,
\left(
\neg r
\,\wedge\,
\LTLG( (\LTLX r) \leftrightarrow (r \leftrightarrow \neg p))
\,\wedge\,
\psi'
\right)
\end{equation*}
In this specification, $\psi'$ is a place-holder for additional guarantees, and not defined explicitly. 
Intuitively, the specification requires the system to track (with its signal $r$) whether the number of time steps or \markterm{rounds} in which $p$ was set to $\TRUE$ is even or odd.
The environment may set the $q$ signal to $\TRUE$ only when an odd number of such rounds has been seen. Additionally, we require the environment to set the $q$ signal to true at least once every two rounds.

This specification is unrealizable, because the strict semantics of the implication prevents the system from proactively setting its $r$ signal to $\FALSE$ for two successive rounds (in violation of the guarantees) -- doing so would force the environment to violate either $\LTLG( q \vee \LTLX q)$ or $\LTLG( \neg q \vee r)$. With the non-strict semantics, the system could in this manner make sure that the overall specification is satisfied (as the assumptions are violated). 
The use of additional signals as above to encode more complex properties is common practice in GR(1) synthesis. Verifying how the different semantics affect the \markterm{realizability} result is a good sanity check for a specification when these more complex properties impose restrictions on the environment behavior.
\begin{example}
Assume that, during the writing process of the specification $\psi$, a new liveness guarantee has just been added to $\psi'$, making the specification unrealizable under the strict implication semantics. This is not an uncommon occurrence in the specification engineering process, as specifications are typically built step-by-step, with assumptions added as needed\cite{DBLP:journals/fmsd/Ehlers12}. As the overall specification $\psi$ is known to be unrealizable due to the liveness guarantees, we would expect $\psi$ to be realizable in the non-strict semantics, as it can set $r$ to erroneous values. If a specification analysis tool now reports that the specification is also unrealizable in the non-strict semantics, then we know that $\psi$ has an error in parts other than $\psi'$. In this way, we could, for example, detect a typo in the first assumption, such as having $\LTLG(q \vee \LTLX p)$ instead of $\LTLG(q \vee \LTLX q)$.
\end{example}
From an implementation perspective, checking whether the strict implication semantics make a difference is not difficult -- we just apply the classical semantics and see if this changes the realizability result. A detailed explanation of how to use the classical implication semantics in GR(1) synthesis was provided in \cite{DBLP:conf/hvc/KleinP10}.

\section{Analyzing the Relationship between Assumptions and Guarantees}
\label{sec:assumptionsAndGuarantees}
In many applications of reactive synthesis, we need to make assumptions about the possible environment behavior in addition to stating the desired system guarantees. This calls for methods to analyze the interaction of the assumptions and guarantees. We consider the following approaches in this context: 

\begin{itemize}
\item Computing statistics about which positions in the \newterm{synthesis game}  are winning/losing for the system
\item Computing information about positions from which the system can falsify the (safety) assumptions
\item Detecting superfluous assumptions
\item Analyzing the achievable levels of error-resilience against violations of the environment assumptions
\end{itemize}
The techniques in this section focus on analyzing a specification: in a debugging context, the analysis results are meant to be compared against the expectations of the system designer. Results of such analyses often indicate potential problems and justify a closer look, as demonstrated by the examples in this section.

\subsection{Winning and Losing Positions in Synthesis Games}
The initialization assumptions and guarantees in GR(1) specifications ensure that the interaction between the environment and the system has a well-defined starting condition.
Depending on the specification and the amount of \emph{\markterm{pre-synthesis}} performed to encode it in GR(1) form, the number of admissible starting positions that are winning for the system may be quite large or relatively small. By giving the specification engineer the ability to check the true number of these winning states against the number expected, we provide an easy-to-use sanity-check.

For a more fine-grained analysis, we distinguish between positions that satisfy the initialization assumptions and and those that satisfy the initialization guarantees. Additionally, certain \markterm{cubes} of winning/non-winning positions are also useful for sanity checking -- if positions are found to be losing that should not be, then this indicates a problem with the specification, even if the synthesized solution seems to behave in a reasonable manner.

\begin{example}
Consider a simple mutual exclusion protocol with grant pre-an\-noun\-ce\-ment. The input signals to the system are $\AP^I = \{r_1, r_2\}$, the output signals are $\AP^O = \{g_1, g_2, \allowbreak \mathit{promise}_1, \allowbreak \mathit{promise}_2\}$, and we are given the following specification:
\begin{eqnarray*}
\left(
\TRUE
\right)
& \!\!\! \rightarrow_S \!\!\! &  
\big( \neg g_1 \wedge \neg g_2 \wedge
\LTLG ( \neg \LTLX \mathit{promise}_1 \vee \neg \LTLX \mathit{promise}_2 )
\wedge \LTLG (\mathit{promise}_1 \leftrightarrow \LTLX g_1)
\wedge \LTLG (\mathit{promise}_2 \leftrightarrow \LTLX g_2) \\
& & 
{\color{white} (} \wedge \LTLG \LTLF (r_1 \rightarrow \mathit{promise}_1)
\wedge \LTLG \LTLF (r_2 \rightarrow \mathit{promise}_2)
\big)
\end{eqnarray*}
Intuitively, the system is asked to use the \markterm{signals} $\mathit{promise}_1$ and $\mathit{promise}_2$ to pre-announce grants $g_1$ and $g_2$. The requirement that two grants are never given at the same time is implemented by preventing the system from taking a transition in which $\mathit{promise}_1$ and $\mathit{promise}_2$ are $\TRUE$ at the same time, and requiring the system to give precisely the grants that have been pre-announced.

However, an analysis of which positions are winning reveals that there are no losing positions that violate the initialization guarantees, not even the ones in which $\mathit{promise}_1$ and $\mathit{promise}_1$ are both set to $\TRUE$. This indicates that the $\LTLX$ operators in the third guarantee of the specification should be removed in order to prevent the system from giving two grants at the same time in the second time step. Alternatively, we could add $\neg \mathit{promise}_1 \wedge \neg \mathit{promise}_2$ as an initialization guarantee. Then, the system has only one initial variable valuation to choose from, and this would be reflected in the numbers of initial positions that satisfy the initialization guarantees. If we instead change the third guarantee and run the analysis again, we find that the system losing positions are precisely those in which $\mathit{promise}_1$ and $\mathit{promise}_2$ are both set to $\TRUE$. Thus, we could confirm that by this change, the only remaining positions that are not winning for the system are those we expect.
\end{example}

Counting the number of winning positions is easy once a BDD for the set of winning positions has been computed. For computing cubes of winning/non-winning positions, the naive approach is to search for short paths to the $\TRUE$ sink in the BDD. However, this can yield unnecessarily small cubes (i.e., with unnecessarily many \markterm{literals}). Instead, we use the \newterm{implicit cube enumeration} technique on BDDs proposed by Coudert and Madre \cite{DBLP:conf/dac/CoudertM92} in order to compute a so-called \emph{meta-product} BDD. We can then enumerate the largest possible cubes by enumerating paths in the meta-product BDD in which as many variables as possible are mapped to \emph{don't care}.

\subsection{Falsifying the (Safety) Assumptions}

Positions in the synthesis game can be \markterm{winning} for several reasons: either the system can drive the environment into falsifying the assumptions (i.e., ensure that the assumptions are not fulfilled along any trace beginning from that position), or the system can ensure the satisfaction of its guarantees along some traces starting from that position, including all traces on which the assumptions are satisfied.

In general, the desired case is that all assumptions and guarantees are fulfilled on a trace. Yet, for realizable specifications, there exists a possibility that the system enforces an assumption violation. Depending on the application, this may be an intended outcome (e.g., after the environment provides an input that is not explicitly disallowed by the assumptions, but is not possible in the environment in which the system is intended to operate). By giving the user the ability to inspect such cases, we provide a debugging tool for analyzing the interplay of assumptions and guarantees in a specification.

One way of providing this capability is by adding $\LTLG \LTLF(\FALSE)$ to the guarantees and applying the analysis described in the previous subsection. Due to the strict implication semantics in GR(1) synthesis, all positions that remain winning for the system can only be winning by violating some environment assumption.

\begin{example}
\label{example:robot1}
Let us consider the simple two-robot scenario depicted in Figure \ref{fig:twoRobotsFigure}. We want to synthesize a high-level controller for the primary robot, whose task is to cycle between the lower-left and lower-right cells. Both robots can move by one cell in the $x$- and $y$-direction at every step. Furthermore, in order to avoid collisions, we assume that the secondary robot (whose location is an input to the controller) can never choose the primary robot's current position as its next position, and require that the primary robot (whose location is updated by the controller) may not choose a next location that is the same as the secondary robot's next location. 
Lastly, we add the assumption that when the robots are adjacently located, the secondary robot will eventually change its location if the primary one does not -- this prevents livelock.

The above specification is found to be realizable, and all positions are winning (or are labeled by some robot location that is not within the workspace boundaries)\footnote{Note that in the specification, we only require that the robots not perform location changes that lead to collisions. There is no requirement that transitions do not start from colliding positions -- this would lead to fewer winning positions.}.
We find that there are some positions from which the system can enforce an assumption violation. For example, if the secondary robot is in cells 1, 2, or 3, and the primary robot is one cell right of the secondary one, then the system can enforce the falsification of the assumptions. In particular, the secondary robot cannot avoid getting stuck, violating the liveness assumptions. 
This is problematic, as synthesized controllers can exploit this fact (e.g., by letting the primary robot move to cell 4 when the secondary robot is in cell 1 and the primary robot is in cell 5), resulting in undesired system behavior. To prevent this from happening, we can alter the liveness assumption to also be satisfied whenever the robot is in the top-most row. Applying our analysis shows that doing so causes the specification to remain realizable, but there is now no position in the synthesis game from which an assumption violation can be enforced.
\end{example}

\begin{figure}
\centering
\begin{minipage}{0.47\textwidth}
\centering\begin{tikzpicture}[scale=0.7]
\path[fill=green!70!gray] (0,0) rectangle +(1,1);
\path[fill=green!70!gray] (7,0) rectangle +(1,1);
\draw[thick,color=gray] (0, 0) grid (8, 5);
\draw[color=gray,fill=black] (0,3) rectangle (4,4);
\draw[color=gray,fill=black] (4,0) rectangle (5,1);
\draw[fill=blue!60!gray] (0.5,0.5) circle (0.35cm);
\draw[fill=red!80!blue!30!gray] (7.5,4.5) circle (0.30cm);
\node[color=black!80!white,anchor=south east, inner sep=1pt] at (1,4) {\scriptsize \textsf{1}};
\node[color=black!80!white,anchor=south east, inner sep=1pt] at (2,4) {\scriptsize \textsf{2}};
\node[color=black!80!white,anchor=south east, inner sep=1pt] at (3,4) {\scriptsize \textsf{3}};
\node[color=black!80!white,anchor=south east, inner sep=1pt] at (4,4) {\scriptsize \textsf{4}};
\node[color=black!80!white,anchor=south east, inner sep=1pt] at (5,2) {\scriptsize \textsf{5}};
\end{tikzpicture}
\caption{Robot workspace for Example~\ref{example:robot1}. The primary robot starts in the lower left corner of the workspace, while the secondary robot starts in the upper right corner.}
\label{fig:twoRobotsFigure}
\end{minipage}\hfill
\begin{minipage}{0.47\textwidth}
\centering\begin{tikzpicture}[scale=0.58333]
\path[fill=green!70!gray] (0,0) rectangle +(1,1);
\path[fill=green!70!gray] (7,0) rectangle +(1,1);
\draw[thick,color=gray] (0, 0) grid (8, 6);

\draw[color=gray,fill=black] (1,2) -- (1,5) -- (2,5) -- (2,2) -- cycle;
\draw[color=gray,fill=black] (7,5) -- (7,3) -- (6,3) -- (6,4) -- (3,4) -- (3,5) -- cycle;
\draw[color=gray,fill=black] (3,2) -- (3,3) -- (4,3) -- (4,2) -- (5,2) -- (6,2) -- (6,1) -- (3,1) -- cycle;

\draw[color=gray,fill=blue!50!white] (4,0) rectangle (5,1);
\draw[color=gray,fill=blue!50!white] (3,5) rectangle (4,6);

\draw[fill=blue!60!gray] (0.5,0.5) circle (0.35cm);
\end{tikzpicture}
\caption{Robot workspace for Example~\ref{example:robotAndDoors}. Two environment liveness assumptions are added to the specification to ensure that the doors are open infinitely often.}
\label{fig:anotherRobotWorkspace}
\end{minipage}
\end{figure}

\subsection{Detecting Superfluous Assumptions}

When a specification engineer adds an assumption for synthesis, it is typically either to give the system more flexibility in achieving its objectives, or to make the specification realizable in the first place.
However, it can occur that some assumptions are not actually needed; this can happen if, for example, the other assumptions are already strong enough to ensure realizability, or there is already an error in one of the other assumptions.

Detecting such assumption superfluity is easy: we can remove each assumption and check if the specification stays realizable. 
However, such an analysis is coarse-grained, and masks several interesting properties. For example, some assumptions can make more positions in the game winning (which helps to compute a reasonable controller that also works under certain assumption violations \cite{EhlersTopcuHSCC2014}). Some can also simplify strategies in the synthesis game, leading to better controllers in certain applications.
As a remedy, we propose to test each assumption for whether 
\begin{itemize}
\item[(a)] it changes the realizability of a specification,
\item[(b)] it makes more positions in the game winning for the system,
\item[(c)] it reduces the \newterm{reactive distance} (see Section~\ref{secWithDef:reactiveDistance}) from some position to some goal, or
\item[(d)] it reduces the \newterm{reactive distance} to the next goal from some position that is reachable in some implementation of a GR(1) strategy.
\end{itemize}
If none of these conditions hold, then we classify an assumption as superfluous. Otherwise, the results of these tests are supplied to the specification engineer.

\begin{example}
\label{example:robotAndDoors}
Consider the robot workspace given in Figure \ref{fig:anotherRobotWorkspace}. The setting is similar to the one in Example~\ref{example:robot1}, with the exception that there is no secondary robot. Rather, there are two cells that serve as \emph{doors}. The primary robot is not allowed to enter these cells when they are closed, but they are guaranteed to be open in infinitely many time steps (rounds of the synthesis game): for each door, there is a liveness assumption stating that it will be open infinitely often.

Analyzing the example shows that all positions are winning for the system, and the specification stays realizable without the two liveness assumptions. The liveness assumption for the top door is superfluous when the system is trying to make progress towards the lower right corner, but sometimes helps for reaching the lower left corner. However, in \markterm{any implementation that a GR(1) synthesizer would compute}, this assumption is superfluous. We also find that the assumption that the bottom door is always eventually open is useful for both \markterm{system goals}.

From this result, the system engineer may either realize that a part of the specification that makes use of the door at the top has been forgotten, or find that the door at the top \emph{is} useless, and the assumption can be removed.
\end{example}

\subsection{Analyzing the Achievable Levels of Error-resilience Against Violations of the Environment Assumptions}
\label{sec:errorResilience}

Even if a (safety) assumption in a specification is not superfluous, it is sometimes possible to synthesize a system that can tolerate a few \newterm{glitches}, i.e., temporary violations. This is especially true for safety assumptions that are only needed for the system to satisfy some liveness guarantee.

\begin{figure}
\centering\begin{tikzpicture}[scale=0.58333]
\path[pattern=north west lines, pattern color=green!70!gray] (2,3) rectangle +(1,1);
\draw[thick,color=gray] (0, 0) grid (10, 6);

\draw[color=gray,fill=black] (0,2) -- (0,6) -- (5,6) -- (5,2) -- (4,2) -- (4,5) -- (1,5) -- (1,2) -- cycle;

\draw[fill=blue!60!gray] (9.5,5.5) circle (0.35cm);
\end{tikzpicture}
\caption{Robot workspace for Example~\ref{example:errorResilience}.}
\label{fig:robotWorkspaceErrorResilience}
\end{figure}
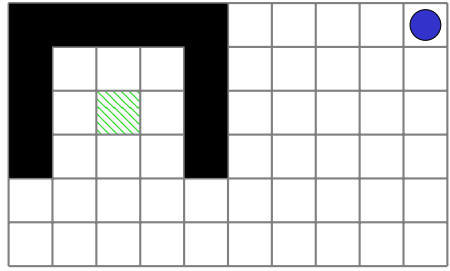

\begin{example}
\label{example:errorResilience}
Consider the robot delivery problem depicted in Figure~\ref{fig:robotWorkspaceErrorResilience}. The current position of the robot is an \emph{input} to the robot controller, and the robot can choose to move left, up, down, or right. Whenever the robot obtains a ``$\mathit{moveit}$'' signal, is must eventually visit the striped region, and must always avoid collisions with the obstacle (collisions with workspace boundaries do not need to be avoided in this specification).

We want to synthesize a robot controller under the following assumptions: 
\begin{enumerate}
\item The $x$ coordinate is always updated according to the robot's requests
\item The $y$ coordinate is always updated according to the robot's requests
\item The robot's position never jumps by more than one cell at a time (in each of the $x$ and $y$ coordinates).
\item The ``$\mathit{moveit}$'' signal may only be issued if the robot controller set the ``$\mathit{ready}$'' output signal to $\TRUE$ in the previous round.
\end{enumerate}
The specification is realizable, as one would guess. Furthermore, the expectation that the robot can tolerate one violation of one of the first two assumptions is met. However, an automated analysis shows that actually, $5$ violations can be tolerated. This hints to the specification engineer that the specification might contain errors. In fact, it is missing the guarantee that the $\mathit{ready}$ signal be set infinitely often. Adding this guarantee yields the expected level of error-resilience, namely that one glitch can be tolerated, as the goal region is one cell away from the obstacle.
\end{example}
The algorithm for computing the number of glitches tolerable by some implementation of a GR(1) specification is described in \cite{EhlersTopcuHSCC2014}.

\section{Input/Output Proposition Analysis}
\label{sec:weakeningAndStrengthening}
In this section, we present techniques to analyze the effects of individual atomic propositions on the realizability of a specification. We first present an approach for analyzing the effect of shuffling the order in which the atomic propositions are set, i.e., deviations from the strict input-first-then-output semantics of system execution. Then we describe how determining the effect of stuck-at-0 or stuck-at-1 faults in the input or output signals can help with specification debugging.

\subsection{Changing the Signal Order}
Recall that in GR(1) synthesis, a two-player game is used to decide realizability and to compute a finite-state automaton implementing the specification. The two players take turns moving in accordance with the assumptions and guarantees respectively, and a winning strategy for the second player (also known as the system player) yields the desired finite-state machine. The first player is usually called the environment player, and has to ensure that the assumptions declared in the specification are met.

There is an implicit turn-taking semantics to the game. At each turn, the environment player moves first, providing a truth assignment to $\AP_I$ for that time step; this assignment must satisfy the assumptions. The system player gets to observe this assignment before providing a truth assignment to $\AP_O$, satisfying the guarantees. This turn-based game yields infinite executions, all of which should satisfy the specification. However, sometimes it might be desirable for the system to move without having observed the environment. In the robotics domain, for example, where propositions in $\AP_O$ correspond to actions such as motion that take non-trivial amounts of time, it may be desirable to move as soon as possible. This motivates us to check at synthesis time which outputs in $\AP_O$ can be assigned truth values at each time step in a fashion that is agnostic to the environment player's choice with respect to $\AP_I$ at that time step. The synthesized machine is thus no longer strictly a Mealy machine -- some outputs may be fixed before the inputs are read.

\begin{example}
Let us consider a robotics example in which the robot is trying to patrol five regions $r_1-r_5$ in some workspace. The regions are connected in a ring, i.e. $r_1$ is connected to $r_2$, $r_2$ to $r_3$ and so on, and $r_5$ is connected to $r_1$. The robot also has a sensor that can sense the presence or absence of a person. Finally, it has a camera, which it can turn on and off. The robot starts in $r_1$, and is supposed to patrol the five regions. It must also turn its camera on if and only if it is sensing a person. The input signals to the robot system are $\AP_I = \{\mathit{person}\}$, and the output signals are $\AP_O = \{r_1, r_2, r_3, r_4, r_5, \allowbreak \mathit{person}\}$, and we have the following specification:
\begin{eqnarray*}
\left(
\TRUE
\right)
& \rightarrow_S & 
\big(
r_1 \wedge 
\LTLG (\LTLX \mathit{camera} \leftrightarrow \LTLX \mathit{person} ) \wedge \bigwedge_{r_i \neq r_j} \LTLG (\neg r_i \vee \neg r_j)\\
&&{\color{white} (}\wedge \LTLG (r_1 \rightarrow \LTLX (r_5 \vee r_1 \vee r_2))\\
&&{\color{white} (}\wedge \LTLG (r_2 \rightarrow \LTLX (r_1 \vee r_2 \vee r_3))\\
&&{\color{white} (}\wedge \LTLG (r_3 \rightarrow \LTLX (r_2 \vee r_3 \vee r_4))\\
&&{\color{white} (}\wedge \LTLG (r_4 \rightarrow \LTLX (r_3 \vee r_4 \vee r_5))\\
&&{\color{white} (}\wedge \LTLG (r_5 \rightarrow \LTLX (r_4 \vee r_5 \vee r_1))\\
&&{\color{white} (}\wedge \LTLG \LTLF (r_1) \wedge \LTLG \LTLF (r_2) \wedge \LTLG \LTLF (r_3) \wedge\LTLG \LTLF (r_4) \wedge \LTLG \LTLF (r_5)
\big)
\end{eqnarray*}
When we analyze the signals to try and move decisions about the outputs to before the inputs have been seen, we find that we can move the decisions about the region signals $\{r_1, r_2, r_3, r_4, r_5\}$, but not the camera. This is because we need to know if we are sensing a person in order to decide whether to turn on the camera, but deciding what motion to take does not require knowledge about the environment signal (i.e. $\mathit{person}$).
This is an expected consequence of the specification, and indicates that this aspect of the specification has been formalized without introducing any errors.
\end{example}

\subsection{Stuck-at-0 and Stuck-at-1 Faults}
\label{sec:stuckFaults}
When testing hardware circuits, the most commonly found problems are stuck-at-0 and stuck-at-1 faults, where some line of the circuit is rendered unable to represent a value other than $\FALSE$ or $\TRUE$, respectively. In the context of synthesis, we can use this concept to check whether, for a realizable specification, there is some easy way for the system to satisfy the specification without setting an output signal in a reactive fashion. Likewise, for unrealizable specifications, we can check whether a specification stays unrealizable if an input signal valuation is fixed. This provides information about the cause of unrealizability of a specification, especially when unrealizability is not completely due to the safety assumptions and guarantees (and thus the approach in Section~\ref{sec:conciseStrategy} cannot be applied).

\begin{example}
Reconsider the setting from Example~\ref{example:errorResilience}. An analysis of the (original, unfixed) specification reveals that the specification is realizable even if the system is also required to set any of the propositions ``$\mathit{up}$'', ``$\mathit{down}$'', ``$\mathit{left}$'', ``$\mathit{right}$'', or ``$\mathit{ready}$'' to $\FALSE $ permanently. This suggests to the specification engineer that the specification is incomplete.
\end{example}

\section{Computing Concise Strategies and Traces}
\label{sec:strategiesAndTraces}

The specification debugging techniques discussed so far are concerned with collecting statistics on the specification and properties of the generated implementation.
It is also interesting to see how a system behaves at runtime. Typically, simulation is applied for this purpose. However, depending on the application, it can be difficult to provide input to the synthesized system that leads to traces that are insightful for the system engineer. To mitigate this problem, we present two approaches for automatically computing artifacts that explain how a system interacts with its environment. In Section~\ref{sec:conciseTrace}, we present an approach to computing a nominal-case behavior trace of the synthesized system. In Section~\ref{sec:conciseStrategy}, we compute an abstract trace-like strategy for the system in case it can enforce a safety assumption violation, or a trace-like strategy for the environment in case it can enforce a safety guarantee violation.

\subsection{Nominal-Case Trace Computation}
\label{sec:conciseTrace}

The key advantage of allowing assumptions in specifications is that it enables the system engineer to restrict the set of environments in which an implementation must work correctly, to those that are of relevance for the application at hand. Thus, in order to observe if a specification is already complete, it makes sense to look at the behavior of the system when the assumptions are satisfied.

However, when simulating a system in order to check whether it already exhibits the intended behavior, it is a nontrivial task to provide inputs that satisfy all assumptions. This is especially true when there are a large number of input propositions with many constraints on them.

To remedy this, we present an approach for finding a trace of the system that exhibits nominal-case behavior. We would call such a trace \newterm{informative} if that term had not already been taken\footnote{In runtime verification, a prefix of a trace is called \emph{informative} for a linear-time specification if a proof of any extension of the prefix to satisfy or violate the specification can be conducted just over the syntactical structure of the specification formula, without quantifying over the possible suffix traces.}.
To obtain such a trace, we first restrict the behavior of the environment such that it only makes choices that are consistent with the safety assumptions, and makes progress towards satisfying the liveness assumptions. In order to compute these choices (i.e. transitions), we inspect the game structure built during GR(1) synthesis \cite{DBLP:journals/jcss/BloemJPPS12}. We treat this game as a \newterm{B\"uchi game} in which it is the aim of the environment to satisfy the liveness assumptions. We then compute a non-deterministic winning strategy for this B\"uchi game. It is non-deterministic because whenever there are multiple next moves that are equally good in terms of the distance to the environment goal, both moves are part of the strategy.

The environment then moves using this non-deterministic strategy, and the system does the same from the strategy computed during GR(1) synthesis. For efficient specification debugging, we augment the trace by information about which goal the environment and system are trying to make progress towards along each transition.

\subsection{Abstract Strategies and Counter-Strategies}
\label{sec:conciseStrategy}
A specification that contains both safety and liveness parts can be rendered realizable or unrealizable due to the safety assumptions and guarantees alone. 
In such a case, either the environment or the system can \emph{win the synthesis game} in a finite number of steps. 
Analyzing such a case can be difficult -- even if a simulation environment is available, it is difficult to trigger the events that are informative to the system engineer. 

We propose the following approach to counter this problem. We compute an \emph{abstract strategy} for the winning player (system or environment) in the synthesis game. The strategy is a mapping from the step number in the game and the atomic proposition controlled by the winning player to either the value of the proposition in the respective round, or to a valuation that depends on the other player's choices.
When computing a strategy for the system, we only need to consider the next choice of input proposition valuations that do not violate the safety assumptions.

Our abstract strategies are uncomparable to the outcomes of previous approaches to counter-strategy generation (such as \cite{DBLP:journals/sttt/KonighoferHB13}), as we represent a strategy in an abstract way instead of introducing branching in the strategy whenever needed. In many cases, this helps to pinpoint the reason for the realizability or unrealizability of a specification. Computing the abstract strategy is performed by repeatedly solving the synthesis game, while checking if constraining the behavior of the winning player in individual rounds changes the outcome.

\begin{example}
\label{example:abstractCounterStrategy}
Consider a robot moving on a grid that has to be in some part of the workspace every $4$ times the input proposition $r$ is set to $\TRUE$. The system is asked to update a counter that keeps track of the number of $r$ requests (modulo $4$) so far.  The signal $r$ may be set to $\TRUE$ only in every second step.

Consider what happens when the specification guarantee that requires the counter to be updated correctly is written incorrectly, such that it does not actually count modulo $4$, allowing the counter to overflow. This means that the environment has a strategy to force the system to violate its safety guarantees. In a complex specification, this type of problem would be hard to observe. However, the abstract counter-strategy given in Table~\ref{tab:abstractCounterStrategy}, which we can compute automatically from the specification, demonstrates the problem quite clearly.
\end{example}

\begin{table}
\small
\renewcommand{\arraystretch}{0.99} 
\newcommand{\itdepends}{\star}
\newcommand{\theendofthetrace}{\text{\sffamily X}}
\centering\begin{tabular}{l||c|c|c|c|c|c|c|c}
\textbf{Atomic proposition / Round} & 0 & 1 & 2 & 3 & 4 & 5 & 6 & 7 \\ \hline \hline
$r$ & $\TRUE$ & $\FALSE$ & $\TRUE$ & $\FALSE$ & $\TRUE$ & $\FALSE$ & $\TRUE$ & $\theendofthetrace$ \\ \hline \hline
$\mathit{counter}$ & $0$ & $1$ & $1$ & $2$ & $2$& $3$& $3$& $\theendofthetrace$ \\ \hline
$\mathit{x}$ & $0$ & $\itdepends$ & $\itdepends$ & $\itdepends$ & $\itdepends$ & $\itdepends$ & $\itdepends$ & $\theendofthetrace$ \\ \hline
$\mathit{y}$ & $0$ & $\itdepends$ & $\itdepends$ & $\itdepends$ & $\itdepends$ & $\itdepends$ & $\itdepends$ & $\theendofthetrace$ 
\end{tabular}
\vspace{-2mm}
\caption{An abstract counter-strategy for Example~\ref{example:abstractCounterStrategy}. The values that depend on the inputs are shown as $\itdepends$, the special value $\theendofthetrace$ indicates that the previous transitions led to a violation of the safety guarantees.}
\label{tab:abstractCounterStrategy}
\end{table}

\section{Conclusion}

In this paper, we presented a report-based specification-debugging approach for reactive synthesis. The core idea is to perform a sequence of analysis steps that allow the system engineers to ``sanity check'' various properties of the specification. Unexpected results indicate wrong or missing parts of the specification or, in some circumstances, potential for strengthening the specification in order to obtain a more desirable reactive controller. 

In addition to the fact that report-based specification debugging offers a push-button approach to specification analysis, it allowed us to employ techniques with a high computational cost, which are only feasible without real-time user interaction (for example, by running over-night). In particular, the techniques in Sections~\ref{sec:errorResilience}, \ref{sec:stuckFaults}, and \ref{sec:conciseStrategy} require many calls to the synthesis engine, each of which takes approximately as much time as realizability checking the original specification. Nevertheless, if a specification is compact enough to allow reactive synthesis, then solving a sequence of synthesis problems of a similar complexity will typically also be feasible.

Our techniques are available as part of the generalized reactivity(1) synthesis tool \textsc{slugs}, which is available for download at \url{https://github.com/ltlmop/slugs}.

The set of techniques presented here is certainly not comprehensive, and many further variations and new ideas are conceivable. Several of the ideas presented in this paper are also adapted to GR(1) specifications, and may need to be generalized before they can be applied to other specification languages. It should be mentioned that the strict structure of a GR(1) specification actually helps with debugging a specification, as it ensures that questions such as the ones in Section~\ref{sec:assumptionsAndGuarantees} are actually well-defined. For example, in general LTL synthesis, it would not make much sense to consider different starting positions in a synthesis game, as the compilation from logic to automata during general LTL synthesis strips them of any meaning with respect to the original specification, unlike in GR(1) synthesis.

\paragraph{Acknowledgements} V. Raman is supported by TerraSwarm, one of six centers of STARnet, a Semiconductor Research Corporation program sponsored by MARCO and DARPA.

\bibliographystyle{eptcs}
\bibliography{bib}

\begin{thebibliography}{10}
\providecommand{\bibitemdeclare}[2]{}
\providecommand{\surnamestart}{}
\providecommand{\surnameend}{}
\providecommand{\urlprefix}{Available at }
\providecommand{\url}[1]{\texttt{#1}}
\providecommand{\href}[2]{\texttt{#2}}
\providecommand{\urlalt}[2]{\href{#1}{#2}}
\providecommand{\doi}[1]{doi:\urlalt{http://dx.doi.org/#1}{#1}}
\providecommand{\bibinfo}[2]{#2}

\bibitemdeclare{inproceedings}{DBLP:conf/movep/AmnellBBDDFHJLMPWY00}
\bibitem{DBLP:conf/movep/AmnellBBDDFHJLMPWY00}
\bibinfo{author}{Tobias \surnamestart Amnell\surnameend}, \bibinfo{author}{Gerd
  \surnamestart Behrmann\surnameend}, \bibinfo{author}{Johan \surnamestart
  Bengtsson\surnameend}, \bibinfo{author}{Pedro~R. \surnamestart
  D'Argenio\surnameend}, \bibinfo{author}{Alexandre \surnamestart
  David\surnameend}, \bibinfo{author}{Ansgar \surnamestart Fehnker\surnameend},
  \bibinfo{author}{Thomas \surnamestart Hune\surnameend},
  \bibinfo{author}{Bertrand \surnamestart Jeannet\surnameend},
  \bibinfo{author}{Kim~Guldstrand \surnamestart Larsen\surnameend},
  \bibinfo{author}{M.~Oliver \surnamestart M{\"o}ller\surnameend},
  \bibinfo{author}{Paul \surnamestart Pettersson\surnameend},
  \bibinfo{author}{Carsten \surnamestart Weise\surnameend} \&
  \bibinfo{author}{Wang \surnamestart Yi\surnameend} (\bibinfo{year}{2000}):
  \emph{\bibinfo{title}{UPPAAL - Now, Next, and Future}}.
\newblock In \bibinfo{editor}{Franck \surnamestart Cassez\surnameend},
  \bibinfo{editor}{Claude \surnamestart Jard\surnameend},
  \bibinfo{editor}{Brigitte \surnamestart Rozoy\surnameend} \&
  \bibinfo{editor}{Mark~Dermot \surnamestart Ryan\surnameend}, editors: {\sl
  \bibinfo{booktitle}{MOVEP}}, {\sl \bibinfo{series}{Lecture Notes in Computer
  Science}} \bibinfo{volume}{2067}, \bibinfo{publisher}{Springer}, pp.
  \bibinfo{pages}{99--124}, \doi{10.1007/3-540-45510-8\_4}.

\bibitemdeclare{article}{DBLP:journals/jcss/BloemJPPS12}
\bibitem{DBLP:journals/jcss/BloemJPPS12}
\bibinfo{author}{Roderick \surnamestart Bloem\surnameend},
  \bibinfo{author}{Barbara \surnamestart Jobstmann\surnameend},
  \bibinfo{author}{Nir \surnamestart Piterman\surnameend},
  \bibinfo{author}{Amir \surnamestart Pnueli\surnameend} \&
  \bibinfo{author}{Yaniv \surnamestart Sa'ar\surnameend}
  (\bibinfo{year}{2012}): \emph{\bibinfo{title}{Synthesis of Reactive(1)
  designs}}.
\newblock {\sl \bibinfo{journal}{J. Comput. Syst. Sci.}}
  \bibinfo{volume}{78}(\bibinfo{number}{3}), pp. \bibinfo{pages}{911--938},
  \doi{10.1016/j.jcss.2011.08.007}.

\bibitemdeclare{inproceedings}{Chatterjee08}
\bibitem{Chatterjee08}
\bibinfo{author}{Krishnendu \surnamestart Chatterjee\surnameend},
  \bibinfo{author}{Thomas~A. \surnamestart Henzinger\surnameend} \&
  \bibinfo{author}{Barbara \surnamestart Jobstmann\surnameend}
  (\bibinfo{year}{{2008}}): \emph{\bibinfo{title}{{Environment Assumptions for
  Synthesis}}}.
\newblock In: {\sl \bibinfo{booktitle}{{International Conference on Concurrency
  Theory (CONCUR)}}}, \bibinfo{publisher}{{Springer-Verlag}},
  \bibinfo{address}{{Berlin, Heidelberg}}, pp. \bibinfo{pages}{{147--161}},
  \doi{10.1007/978-3-540-85361-9\_14}.

\bibitemdeclare{inproceedings}{DBLP:conf/cav/CimattiCGGPRST02}
\bibitem{DBLP:conf/cav/CimattiCGGPRST02}
\bibinfo{author}{Alessandro \surnamestart Cimatti\surnameend},
  \bibinfo{author}{Edmund~M. \surnamestart Clarke\surnameend},
  \bibinfo{author}{Enrico \surnamestart Giunchiglia\surnameend},
  \bibinfo{author}{Fausto \surnamestart Giunchiglia\surnameend},
  \bibinfo{author}{Marco \surnamestart Pistore\surnameend},
  \bibinfo{author}{Marco \surnamestart Roveri\surnameend},
  \bibinfo{author}{Roberto \surnamestart Sebastiani\surnameend} \&
  \bibinfo{author}{Armando \surnamestart Tacchella\surnameend}
  (\bibinfo{year}{2002}): \emph{\bibinfo{title}{NuSMV 2: An OpenSource Tool for
  Symbolic Model Checking}}.
\newblock In \bibinfo{editor}{Ed~\surnamestart Brinksma\surnameend} \&
  \bibinfo{editor}{Kim~Guldstrand \surnamestart Larsen\surnameend}, editors:
  {\sl \bibinfo{booktitle}{CAV}}, {\sl \bibinfo{series}{Lecture Notes in
  Computer Science}} \bibinfo{volume}{2404}, \bibinfo{publisher}{Springer}, pp.
  \bibinfo{pages}{359--364}, \doi{10.1007/3-540-45657-0\_29}.

\bibitemdeclare{inproceedings}{Cimatti08}
\bibitem{Cimatti08}
\bibinfo{author}{Alessandro \surnamestart Cimatti\surnameend},
  \bibinfo{author}{Marco \surnamestart Roveri\surnameend},
  \bibinfo{author}{Viktor \surnamestart Schuppan\surnameend} \&
  \bibinfo{author}{Andrei \surnamestart Tchaltsev\surnameend}
  (\bibinfo{year}{{2008}}): \emph{\bibinfo{title}{{Diagnostic Information for
  Realizability}}}.
\newblock In: {\sl \bibinfo{booktitle}{{Verification, Model Checking, and
  Abstract Interpretation (VMCAI)}}}, pp. \bibinfo{pages}{{52--67}},
  \doi{10.1007/978-3-540-78163-9\_9}.

\bibitemdeclare{inproceedings}{Cimatti07}
\bibitem{Cimatti07}
\bibinfo{author}{Alessandro \surnamestart Cimatti\surnameend},
  \bibinfo{author}{Marco \surnamestart Roveri\surnameend},
  \bibinfo{author}{Viktor \surnamestart Schuppan\surnameend} \&
  \bibinfo{author}{Stefano \surnamestart Tonetta\surnameend}
  (\bibinfo{year}{{2007}}): \emph{\bibinfo{title}{{Boolean Abstraction for
  Temporal Logic Satisfiability}}}.
\newblock In: {\sl \bibinfo{booktitle}{{Computer Aided Verification (CAV)}}},
  pp. \bibinfo{pages}{{532--546}}, \doi{10.1007/978-3-540-73368-3\_53}.

\bibitemdeclare{inproceedings}{DBLP:conf/dac/CoudertM92}
\bibitem{DBLP:conf/dac/CoudertM92}
\bibinfo{author}{Olivier \surnamestart Coudert\surnameend} \&
  \bibinfo{author}{Jean~Christophe \surnamestart Madre\surnameend}
  (\bibinfo{year}{1992}): \emph{\bibinfo{title}{Implicit and Incremental
  Computation of Primes and Essential Primes of Boolean Functions}}.
\newblock In: {\sl \bibinfo{booktitle}{DAC}}, pp. \bibinfo{pages}{36--39}.
\newblock \urlprefix\url{http://portal.acm.org/citation.cfm?id=113938.113929}.

\bibitemdeclare{article}{DBLP:journals/fmsd/Ehlers12}
\bibitem{DBLP:journals/fmsd/Ehlers12}
\bibinfo{author}{R{\"u}diger \surnamestart Ehlers\surnameend}
  (\bibinfo{year}{2012}): \emph{\bibinfo{title}{Symbolic bounded synthesis}}.
\newblock {\sl \bibinfo{journal}{Formal Methods in System Design}}
  \bibinfo{volume}{40}(\bibinfo{number}{2}), pp. \bibinfo{pages}{232--262},
  \doi{10.1007/s10703-011-0137-x}.

\bibitemdeclare{phdthesis}{Ehlers2013}
\bibitem{Ehlers2013}
\bibinfo{author}{R{\"u}diger \surnamestart Ehlers\surnameend}
  (\bibinfo{year}{2013}): \emph{\bibinfo{title}{Symmetric and Efficient
  Synthesis}}.
\newblock Ph.D. thesis, \bibinfo{school}{Saarland University}.

\bibitemdeclare{inproceedings}{EhlersTopcuHSCC2014}
\bibitem{EhlersTopcuHSCC2014}
\bibinfo{author}{R{\"u}diger \surnamestart Ehlers\surnameend} \&
  \bibinfo{author}{Ufuk \surnamestart Topcu\surnameend} (\bibinfo{year}{2014}):
  \emph{\bibinfo{title}{Resilience to Intermittent Assumption Violations in
  Reactive Synthesis}}.
\newblock In: {\sl \bibinfo{booktitle}{17th International Conference on Hybrid
  Systems: Computation and Control (HSCC)}}, pp. \bibinfo{pages}{203--212},
  \doi{10.1145/2562059.2562128}.

\bibitemdeclare{inproceedings}{DBLP:conf/aaai/FinucaneJK11}
\bibitem{DBLP:conf/aaai/FinucaneJK11}
\bibinfo{author}{Cameron \surnamestart Finucane\surnameend},
  \bibinfo{author}{Gangyuan \surnamestart Jing\surnameend} \&
  \bibinfo{author}{Hadas \surnamestart Kress-Gazit\surnameend}
  (\bibinfo{year}{2011}): \emph{\bibinfo{title}{Designing Reactive Robot
  Controllers with LTLMoP}}.
\newblock In: {\sl \bibinfo{booktitle}{Automated Action Planning for Autonomous
  Mobile Robots}}, {\sl \bibinfo{series}{AAAI Workshops}}
  \bibinfo{volume}{WS-11-09}, \bibinfo{publisher}{AAAI}.
\newblock
  \urlprefix\url{http://www.aaai.org/ocs/index.php/WS/AAAIW11/paper/view/3982}.

\bibitemdeclare{inproceedings}{DBLP:conf/hvc/FismanKSV08}
\bibitem{DBLP:conf/hvc/FismanKSV08}
\bibinfo{author}{Dana \surnamestart Fisman\surnameend}, \bibinfo{author}{Orna
  \surnamestart Kupferman\surnameend}, \bibinfo{author}{Sarai \surnamestart
  Sheinvald-Faragy\surnameend} \& \bibinfo{author}{Moshe~Y. \surnamestart
  Vardi\surnameend} (\bibinfo{year}{2008}): \emph{\bibinfo{title}{A Framework
  for Inherent Vacuity}}.
\newblock In \bibinfo{editor}{Hana \surnamestart Chockler\surnameend} \&
  \bibinfo{editor}{Alan~J. \surnamestart Hu\surnameend}, editors: {\sl
  \bibinfo{booktitle}{Haifa Verification Conference}}, {\sl
  \bibinfo{series}{Lecture Notes in Computer Science}} \bibinfo{volume}{5394},
  \bibinfo{publisher}{Springer}, pp. \bibinfo{pages}{7--22},
  \doi{10.1007/978-3-642-01702-5\_7}.

\bibitemdeclare{inproceedings}{DBLP:conf/hvc/KleinP10}
\bibitem{DBLP:conf/hvc/KleinP10}
\bibinfo{author}{Uri \surnamestart Klein\surnameend} \& \bibinfo{author}{Amir
  \surnamestart Pnueli\surnameend} (\bibinfo{year}{2010}):
  \emph{\bibinfo{title}{Revisiting Synthesis of {GR(1)} Specifications}}.
\newblock In: {\sl \bibinfo{booktitle}{Haifa Verification Conference (HVC)}},
  pp. \bibinfo{pages}{161--181}, \doi{10.1007/978-3-642-19583-9\_16}.

\bibitemdeclare{inproceedings}{DBLP:conf/hvc/KonighoferHB10}
\bibitem{DBLP:conf/hvc/KonighoferHB10}
\bibinfo{author}{Robert \surnamestart K{\"o}nighofer\surnameend},
  \bibinfo{author}{Georg \surnamestart Hofferek\surnameend} \&
  \bibinfo{author}{Roderick \surnamestart Bloem\surnameend}
  (\bibinfo{year}{2010}): \emph{\bibinfo{title}{Debugging Unrealizable
  Specifications with Model-Based Diagnosis}}.
\newblock In: {\sl \bibinfo{booktitle}{Haifa Verification Conference}}, pp.
  \bibinfo{pages}{29--45}, \doi{10.1007/978-3-642-19583-9\_8}.

\bibitemdeclare{article}{DBLP:journals/sttt/KonighoferHB13}
\bibitem{DBLP:journals/sttt/KonighoferHB13}
\bibinfo{author}{Robert \surnamestart K{\"o}nighofer\surnameend},
  \bibinfo{author}{Georg \surnamestart Hofferek\surnameend} \&
  \bibinfo{author}{Roderick \surnamestart Bloem\surnameend}
  (\bibinfo{year}{2013}): \emph{\bibinfo{title}{Debugging formal
  specifications: a practical approach using model-based diagnosis and
  counterstrategies}}.
\newblock {\sl \bibinfo{journal}{STTT}}
  \bibinfo{volume}{15}(\bibinfo{number}{5-6}), pp. \bibinfo{pages}{563--583},
  \doi{10.1007/s10009-011-0221-y}.

\bibitemdeclare{inproceedings}{DBLP:conf/icra/Kress-GazitFP07}
\bibitem{DBLP:conf/icra/Kress-GazitFP07}
\bibinfo{author}{Hadas \surnamestart Kress-Gazit\surnameend},
  \bibinfo{author}{Georgios~E. \surnamestart Fainekos\surnameend} \&
  \bibinfo{author}{George~J. \surnamestart Pappas\surnameend}
  (\bibinfo{year}{2007}): \emph{\bibinfo{title}{Where's Waldo? Sensor-Based
  Temporal Logic Motion Planning}}.
\newblock In: {\sl \bibinfo{booktitle}{ICRA}}, \bibinfo{publisher}{IEEE}, pp.
  \bibinfo{pages}{3116--3121}, \doi{10.1109/ROBOT.2007.363946}.

\bibitemdeclare{article}{KGF}
\bibitem{KGF}
\bibinfo{author}{Hadas \surnamestart Kress-Gazit\surnameend},
  \bibinfo{author}{Georgios~E. \surnamestart Fainekos\surnameend} \&
  \bibinfo{author}{George~J. \surnamestart Pappas\surnameend}
  (\bibinfo{year}{{2009}}): \emph{\bibinfo{title}{Temporal-Logic-Based Reactive
  Mission and Motion Planning}}.
\newblock {\sl \bibinfo{journal}{IEEE Transactions on Robotics}}
  \bibinfo{volume}{{25}}(\bibinfo{number}{{6}}), pp.
  \bibinfo{pages}{{1370--1381}}, \doi{10.1109/TRO.2009.2030225}.

\bibitemdeclare{inproceedings}{Li11}
\bibitem{Li11}
\bibinfo{author}{Wenchao \surnamestart Li\surnameend}, \bibinfo{author}{Lili
  \surnamestart Dworkin\surnameend} \& \bibinfo{author}{Sanjit~A. \surnamestart
  Seshia\surnameend} (\bibinfo{year}{{2011}}): \emph{\bibinfo{title}{{Mining
  Assumptions for Synthesis}}}.
\newblock In: {\sl \bibinfo{booktitle}{{{{ACM-IEEE}} International Conference
  on Formal Methods and Models for Codesign (MEMOCODE)}}}, pp.
  \bibinfo{pages}{{43--50}}, \doi{10.1109/MEMCOD.2011.5970509}.

\bibitemdeclare{article}{DBLP:journals/taosd/MaozS13}
\bibitem{DBLP:journals/taosd/MaozS13}
\bibinfo{author}{Shahar \surnamestart Maoz\surnameend} \&
  \bibinfo{author}{Yaniv \surnamestart Sa'ar\surnameend}
  (\bibinfo{year}{2013}): \emph{\bibinfo{title}{Two-Way Traceability and
  Conflict Debugging for AspectLTL Programs}}.
\newblock {\sl \bibinfo{journal}{T. Aspect-Oriented Software Development}}
  \bibinfo{volume}{10}, pp. \bibinfo{pages}{39--72},
  \doi{10.1007/978-3-642-36964-3\_2}.

\bibitemdeclare{article}{Nuzzo13}
\bibitem{Nuzzo13}
\bibinfo{author}{P.~\surnamestart Nuzzo\surnameend},
  \bibinfo{author}{H.~\surnamestart Xu\surnameend},
  \bibinfo{author}{N.~\surnamestart Ozay\surnameend}, \bibinfo{author}{J.B.
  \surnamestart Finn\surnameend}, \bibinfo{author}{A.L. \surnamestart
  Sangiovanni-Vincentelli\surnameend}, \bibinfo{author}{R.M. \surnamestart
  Murray\surnameend}, \bibinfo{author}{A.~\surnamestart Donze\surnameend} \&
  \bibinfo{author}{S.A. \surnamestart Seshia\surnameend}
  (\bibinfo{year}{2013}): \emph{\bibinfo{title}{A Contract-Based Methodology
  for Aircraft Electric Power System Design}}.
\newblock {\sl \bibinfo{journal}{Access, IEEE}}
  \bibinfo{volume}{PP}(\bibinfo{number}{99}), pp. \bibinfo{pages}{1--1},
  \doi{10.1109/ACCESS.2013.2295764}.

\bibitemdeclare{inproceedings}{DBLP:conf/iccps/OzayTMW11}
\bibitem{DBLP:conf/iccps/OzayTMW11}
\bibinfo{author}{Necmiye \surnamestart Ozay\surnameend}, \bibinfo{author}{Ufuk
  \surnamestart Topcu\surnameend}, \bibinfo{author}{Richard~M. \surnamestart
  Murray\surnameend} \& \bibinfo{author}{Tichakorn \surnamestart
  Wongpiromsarn\surnameend} (\bibinfo{year}{2011}):
  \emph{\bibinfo{title}{Distributed Synthesis of Control Protocols for Smart
  Camera Networks}}.
\newblock In: {\sl \bibinfo{booktitle}{ICCPS}}, \bibinfo{publisher}{IEEE}, pp.
  \bibinfo{pages}{45--54}, \doi{10.1109/ICCPS.2011.22}.

\bibitemdeclare{inproceedings}{DBLP:conf/cav/PeterEM11}
\bibitem{DBLP:conf/cav/PeterEM11}
\bibinfo{author}{Hans-J{\"o}rg \surnamestart Peter\surnameend},
  \bibinfo{author}{R{\"u}diger \surnamestart Ehlers\surnameend} \&
  \bibinfo{author}{Robert \surnamestart Mattm{\"u}ller\surnameend}
  (\bibinfo{year}{2011}): \emph{\bibinfo{title}{Synthia: Verification and
  Synthesis for Timed Automata}}.
\newblock In \bibinfo{editor}{Ganesh \surnamestart Gopalakrishnan\surnameend}
  \& \bibinfo{editor}{Shaz \surnamestart Qadeer\surnameend}, editors: {\sl
  \bibinfo{booktitle}{CAV}}, {\sl \bibinfo{series}{Lecture Notes in Computer
  Science}} \bibinfo{volume}{6806}, \bibinfo{publisher}{Springer}, pp.
  \bibinfo{pages}{649--655}, \doi{10.1007/978-3-642-22110-1\_52}.

\bibitemdeclare{inproceedings}{DBLP:conf/focs/Pnueli77}
\bibitem{DBLP:conf/focs/Pnueli77}
\bibinfo{author}{Amir \surnamestart Pnueli\surnameend} (\bibinfo{year}{1977}):
  \emph{\bibinfo{title}{The Temporal Logic of Programs}}.
\newblock In: {\sl \bibinfo{booktitle}{FOCS}}, \bibinfo{publisher}{IEEE}, pp.
  \bibinfo{pages}{46--57}.

\bibitemdeclare{inproceedings}{DBLP:conf/icalp/PnueliR89}
\bibitem{DBLP:conf/icalp/PnueliR89}
\bibinfo{author}{Amir \surnamestart Pnueli\surnameend} \& \bibinfo{author}{Roni
  \surnamestart Rosner\surnameend} (\bibinfo{year}{1989}):
  \emph{\bibinfo{title}{On the Synthesis of an Asynchronous Reactive Module}}.
\newblock In: {\sl \bibinfo{booktitle}{ICALP}}, pp. \bibinfo{pages}{652--671}.

\bibitemdeclare{inproceedings}{Raman_CAV11}
\bibitem{Raman_CAV11}
\bibinfo{author}{Vasumathi \surnamestart Raman\surnameend} \&
  \bibinfo{author}{Hadas \surnamestart Kress-Gazit\surnameend}
  (\bibinfo{year}{{2011}}): \emph{\bibinfo{title}{{Analyzing Unsynthesizable
  Specifications for High-Level Robot Behavior Using {{LTLMoP}}}}}.
\newblock In: {\sl \bibinfo{booktitle}{{Computer Aided Verification (CAV)}}},
  pp. \bibinfo{pages}{{663--668}}, \doi{10.1007/978-3-642-22110-1\_54}.

\bibitemdeclare{article}{Raman_TRO13}
\bibitem{Raman_TRO13}
\bibinfo{author}{Vasumathi \surnamestart Raman\surnameend} \&
  \bibinfo{author}{Hadas \surnamestart Kress-Gazit\surnameend}
  (\bibinfo{year}{2013}): \emph{\bibinfo{title}{Explaining Impossible
  High-Level Robot Behaviors}}.
\newblock {\sl \bibinfo{journal}{IEEE Transactions on Robotics}}
  \bibinfo{volume}{29}(\bibinfo{number}{1}), pp. \bibinfo{pages}{94--104},
  \doi{10.1109/TRO.2012.2214558}.

\bibitemdeclare{inproceedings}{Raman_IROS13}
\bibitem{Raman_IROS13}
\bibinfo{author}{Vasumathi \surnamestart Raman\surnameend} \&
  \bibinfo{author}{Hadas \surnamestart Kress-Gazit\surnameend}
  (\bibinfo{year}{2013}): \emph{\bibinfo{title}{Towards minimal explanations of
  unsynthesizability for high-level robot behaviors}}.
\newblock In: {\sl \bibinfo{booktitle}{IROS}}, \bibinfo{publisher}{IEEE}, pp.
  \bibinfo{pages}{757--762}, \doi{10.1109/IROS.2013.6696436}.

\bibitemdeclare{inproceedings}{DBLP:conf/icra/RamanPK13}
\bibitem{DBLP:conf/icra/RamanPK13}
\bibinfo{author}{Vasumathi \surnamestart Raman\surnameend},
  \bibinfo{author}{Nir \surnamestart Piterman\surnameend} \&
  \bibinfo{author}{Hadas \surnamestart Kress-Gazit\surnameend}
  (\bibinfo{year}{2013}): \emph{\bibinfo{title}{Provably correct continuous
  control for high-level robot behaviors with actions of arbitrary execution
  durations}}.
\newblock In: {\sl \bibinfo{booktitle}{ICRA}}, \bibinfo{publisher}{IEEE}, pp.
  \bibinfo{pages}{4075--4081}, \doi{10.1109/ICRA.2013.6631152}.

\bibitemdeclare{inproceedings}{Schuppan09}
\bibitem{Schuppan09}
\bibinfo{author}{Viktor \surnamestart Schuppan\surnameend}
  (\bibinfo{year}{2009}): \emph{\bibinfo{title}{Towards a Notion of
  Unsatisfiable Cores for {LTL}}}.
\newblock In: {\sl \bibinfo{booktitle}{Fundamentals of Software Engineering
  ({FSEN})}}, pp. \bibinfo{pages}{129--145},
  \doi{10.1007/978-3-642-11623-0\_7}.

\bibitemdeclare{inproceedings}{Shlyakhter03}
\bibitem{Shlyakhter03}
\bibinfo{author}{I.~\surnamestart Shlyakhter\surnameend},
  \bibinfo{author}{R.~\surnamestart Seater\surnameend},
  \bibinfo{author}{D.~\surnamestart Jackson\surnameend},
  \bibinfo{author}{M.~\surnamestart Sridharan\surnameend} \&
  \bibinfo{author}{M.~\surnamestart Taghdiri\surnameend}
  (\bibinfo{year}{{2003}}): \emph{\bibinfo{title}{{Debugging Overconstrained
  Declarative Models Using Unsatisfiable Cores}}}.
\newblock In: {\sl \bibinfo{booktitle}{{IEEE International Conference on
  Automated Software Engineering (ASE)}}}, pp. \bibinfo{pages}{{94--105}},
  \doi{{10.1109/ASE.2003.1240298}}.

\bibitemdeclare{inproceedings}{DBLP:conf/fmcad/SohailS09}
\bibitem{DBLP:conf/fmcad/SohailS09}
\bibinfo{author}{Saqib \surnamestart Sohail\surnameend} \&
  \bibinfo{author}{Fabio \surnamestart Somenzi\surnameend}
  (\bibinfo{year}{2009}): \emph{\bibinfo{title}{Safety first: A two-stage
  algorithm for {LTL} games}}.
\newblock In \bibinfo{editor}{Armin \surnamestart Biere\surnameend} \&
  \bibinfo{editor}{Carl \surnamestart Pixley\surnameend}, editors: {\sl
  \bibinfo{booktitle}{FMCAD}}, \bibinfo{publisher}{IEEE}, pp.
  \bibinfo{pages}{77--84}, \doi{10.1109/FMCAD.2009.5351138}.

\bibitemdeclare{inproceedings}{Wongpiromsarn2011-infotech}
\bibitem{Wongpiromsarn2011-infotech}
\bibinfo{author}{Tichakorn \surnamestart Wongpiromsarn\surnameend},
  \bibinfo{author}{Ufuk \surnamestart Topcu\surnameend} \&
  \bibinfo{author}{Richard~M. \surnamestart Murray\surnameend}
  (\bibinfo{year}{2011}): \emph{\bibinfo{title}{Formal synthesis of embedded
  control software for vehicle management systems}}.
\newblock In: {\sl \bibinfo{booktitle}{AIAA Infotech@Aerospace}},
  \doi{10.2514/6.2011-1506}.

\bibitemdeclare{article}{Nok10}
\bibitem{Nok10}
\bibinfo{author}{Tichakorn \surnamestart Wongpiromsarn\surnameend},
  \bibinfo{author}{Ufuk \surnamestart Topcu\surnameend} \&
  \bibinfo{author}{Richard~M. \surnamestart Murray\surnameend}
  (\bibinfo{year}{2012}): \emph{\bibinfo{title}{Receding Horizon Temporal Logic
  Planning}}.
\newblock {\sl \bibinfo{journal}{IEEE Trans. Automat. Contr.}}
  \bibinfo{volume}{57}(\bibinfo{number}{11}), pp. \bibinfo{pages}{2817--2830},
  \doi{10.1109/TAC.2012.2195811}.

\bibitemdeclare{inproceedings}{DBLP:conf/hybrid/WongpiromsarnTOXM11}
\bibitem{DBLP:conf/hybrid/WongpiromsarnTOXM11}
\bibinfo{author}{Tichakorn \surnamestart Wongpiromsarn\surnameend},
  \bibinfo{author}{Ufuk \surnamestart Topcu\surnameend},
  \bibinfo{author}{Necmiye \surnamestart Ozay\surnameend},
  \bibinfo{author}{Huan \surnamestart Xu\surnameend} \&
  \bibinfo{author}{Richard~M. \surnamestart Murray\surnameend}
  (\bibinfo{year}{2011}): \emph{\bibinfo{title}{TuLiP: a software toolbox for
  receding horizon temporal logic planning}}.
\newblock In \bibinfo{editor}{Marco \surnamestart Caccamo\surnameend},
  \bibinfo{editor}{Emilio \surnamestart Frazzoli\surnameend} \&
  \bibinfo{editor}{Radu \surnamestart Grosu\surnameend}, editors: {\sl
  \bibinfo{booktitle}{HSCC}}, \bibinfo{publisher}{ACM}, pp.
  \bibinfo{pages}{313--314}, \doi{10.1145/1967701.1967747}.

\end{thebibliography}

\end{document}